\documentclass[conference]{IEEEtran}
\usepackage[nolist]{acronym}
\usepackage{orcidlink}
\hypersetup{hidelinks}
\usepackage{graphicx} 
\usepackage{amsmath}
\usepackage{amsfonts}
\usepackage{float}
\usepackage{xcolor} 
\usepackage{tabularx} 
\usepackage{booktabs}
\usepackage{cite}
\usepackage[absolute,overlay]{textpos}
\newcolumntype{Y}{>{\centering\arraybackslash}X}
\usepackage[utf8]{inputenc}
\usepackage[caption=false,font=small,labelfont=sf,textfont=sf]{subfig} 
\usepackage{tikz}
\usepackage{pgfplots}
\usetikzlibrary{patterns}
\pgfplotsset{compat=newest}
\newlength\fheight
\newlength\fwidth
\usepackage[margin=0.75in, bottom=1.05in]{geometry}
\usepackage[all]{nowidow}
\DeclareMathOperator*{\argmin}{\arg\!\min}

\usepackage{soul,color}

\title{Indirect Estimation of SINR \\ via  SSB and CSI-RS RSRP in 5G NR}
\vspace{-2cm}
\author{
    \IEEEauthorblockN{
        Leonardo Spampinato\IEEEauthorrefmark{1}\orcidlink{0009-0000-2629-8835}, 
        Mahamadou Togola\IEEEauthorrefmark{2},
        Matteo Bernabè\IEEEauthorrefmark{1}\orcidlink{0009-0005-5432-3053},
        Azim Akhtarshenas\IEEEauthorrefmark{1}, \\
        Lorenzo Mario Amorosa\IEEEauthorrefmark{3}\orcidlink{0000-0002-0405-9611},
        David López-Pérez\IEEEauthorrefmark{1}\orcidlink{0000-0003-1605-7418} 
    } \\
    \vspace{-.4cm}
    \IEEEauthorblockA{\IEEEauthorrefmark{1}ITEAM Research Institute, Universitat Politècnica de València, Valencia, Spain; \\ \IEEEauthorrefmark{2}CentraleSupélec, Gif-sur-Yvette, France; \IEEEauthorrefmark{3}DEI Department, University of Bologna, Bologna, Italy;}
    \vspace{-0.8cm}
}

\IEEEoverridecommandlockouts
\IEEEpubid{\makebox[\columnwidth]{979-8-3195-0489-0/26/\$31.00~\copyright2026 IEEE \hfill}
\hspace{\columnsep}\makebox[\columnwidth]{ }}

\acrodef{5G}{fifth-generation}
\acrodef{UAV}{unmanned aerial vehicle}
\acrodef{UABS}{Unmanned aerial base station}
\acrodef{MBS}{macro base station}
\acrodef{GUE}{ground user equipment}
\acrodef{RRM}{radio resource management}
\acrodef{SUMO}{simulation urban mobility}
\acrodef{RL}{reinforcement learning}
\acrodef{TD}{temporal difference}
\acrodef{MEC}{mobile edge computing}
\acrodef{ILP}{integer linear programming}
\acrodef{MDP}{Markov decision process}
\acrodef{UMa}{urban macro}
\acrodef{SNR}{signal-to-noise ratio}
\acrodef{LoS}{line-of-sight}
\acrodef{NLoS}{non-\ac{LoS}}
\acrodef{BS}{base station}
\acrodef{RF}{radio frequency}
\acrodef{3DQN}{dueling double deep Q-network}
\acrodef{V2X}{vehicle-to-everything}
\acrodef{QoS}{quality of service}
\acrodef{DRL}{deep reinforcement learning}
\acrodef{RRA}{radio resource assignment}
\acrodef{LOS}{line-of-sight}
\acrodef{NLOS}{non-\ac{LOS}}
\acrodef{SSB}{synchronization signal block}
\acrodef{CSI-RS}{channel state information-reference signal}
\acrodef{UE}{user equipment}
\acrodef{PRB}{physical resource block}
\acrodef{SINR}{signal-to-interference-plus-noise ratio}
\acrodef{3GPP}{third-generation partnership project}
\acrodef{RSRP}{reference signal received power}
\acrodef{MSE}{mean squared error}
\acrodef{MLP}{multi layer perceptron}
\acrodef{RMSE}{root mean squared error}
\acrodef{NR}{new radio}
\acrodef{3GPP}{third generation partnership project}
\acrodef{UPA}{uniform planar array}
\acrodef{mMIMO}{massive multiple-input multiple-output}

\acrodef{RS}{reference signal}
\acrodef{2D-DFT}{two dimensional discrete Fourier transform}
\acrodef{PSS}{primary synchronization signal}
\acrodef{SSS}{secondary synchronization signal}
\acrodef{PBCH}{physical broadcast channel}
\acrodef{FR1}{frequency range 1}
\acrodef{PMI}{precoding matrix indicator}
\acrodef{ML}{machine learning}

\begin{document}
\maketitle
\IEEEpubidadjcol
\begin{textblock*}{3cm}(17cm,1cm)
\fontsize{10}{12}\selectfont (Special Session)
\end{textblock*}
\begin{abstract}
Predicting \ac{UE} performance is essential for proactive network control, resource management, and digital twin sandboxes. However, the inherent flexibility and complexity of beam-based 5G \ac{NR} networks make accurate performance forecasting highly challenging. This paper proposes a data-driven approach to predict the average downlink \ac{SINR} relying exclusively on standardized reference-signal measurements, namely \ac{SSB} and \ac{CSI-RS} \ac{RSRP}. We formulate this prediction as a supervised learning problem and evaluate various input feature representations using a \ac{3GPP}-compliant synthetic dataset. Our analysis reveals that filtering measurements based on active \ac{CSI-RS} beams significantly enhances prediction accuracy while reducing input dimensionality. This activity-aware strategy demonstrates the strong viability of machine learning models for proactive network optimization.
\end{abstract}

\acresetall
\section{Introduction}
\label{sec:intro}

The densification of \ac{5G} \ac{NR} networks,
together with the adoption of \ac{mMIMO} antenna arrays and beam-based transmissions,
has significantly increased the flexibility of cellular systems.
By steering synchronization, reference, and data signals over multiple spatial directions,
\ac{5G} \ac{NR} networks can improve coverage, spatial reuse, and \ac{UE} throughput.
At the same time, this flexibility makes radio resource management decisions more complex,
since the performance experienced by a \ac{UE} depends not only on the serving cell,
but also on the selected transmission beam, the activity of neighboring cells, and the interference generated over the same time-frequency resources.

A long-term objective for modern radio access networks is to correctly predict the performance that a \ac{UE} would experience under a prospective serving configuration, for example, simulated within a digital twin sandbox,
before such a configuration is actually applied to the real network.
This capability is relevant to several network functions. For example, in mobility management, the network should assess whether a candidate target cell can provide sufficient service quality before triggering a handover. Similarly, in energy-saving operation, a cell should be switched off only if its served \acp{UE} can be adequately supported by neighbouring cells.
In both cases, the key challenge is to estimate the \ac{UE} performance under radio conditions that may not yet be directly observable.
 
For example, the data rate of an \ac{UE} can be derived from the available bandwidth and the data channel \ac{SINR}. While the allocated bandwidth can be derived from cell load and scheduling policies, predicting the data channel \ac{SINR} is very difficult. This complexity arises because such \ac{SINR} depends not only on the local radio channel and the transmit power, but also on the complex interference dynamics of neighbouring cells.
Moreover, this challenge is further amplified in beam-based \ac{5G} \ac{NR} systems.
During initial access,
\acp{UE} measure the \ac{RSRP} associated with \acp{SSB},
which are periodically transmitted to support synchronization and cell selection~\cite{3GPP38211,3GPP38214,DahlmanBook}.
However, \ac{SSB} beams are not the beams used for data transmission.
Data transmission occurs over beams derived from beam refinement and precoding procedures based on \acp{CSI-RS},
which provide a more direct indication of the beam-domain conditions relevant for data scheduling.
Therefore, inferring data-channel \ac{SINR} from reference-signal measurements is non-trivial,
especially because the interference experienced by a \ac{UE} depends on data scheduling and which beams are actively carrying data.

\subsection{Related Work}
Recent works have investigated predicting \acp{UE} \ac{SINR} in cellular networks.
In particular, \ac{ML} has been used to infer \ac{SINR} from historical channel-quality observations, to reduce reference-signal overhead and improve radio resource management~\cite{ref_ullah_sinr_prediction}.
Other works have explored the use of recurrent and tree-based predictors for \ac{SINR} forecasting, leveraging mobility attributes, measured signal quality, and radio environment maps~\cite{ref_mallikarjun_sinr_pcn}.
Related line of work studies the potential of chaining the prediction of reference-signal measurements to guide network control policies, such as handover optimization and mobility management~\cite{ref_lima_handover}, or energy-saving cell switch-off strategies,
where cell on/off decisions must preserve user service quality while reducing network power consumption~\cite{ref_choi_cell_onoff}.
These studies show the potential of data-driven radio-metric prediction for proactive network control.
However, existing approaches typically focus on temporal forecasting, coverage-map construction, handover triggering, or cell on/off optimization,
and do not generally differentiate channel types.

\subsection{Contribution}
In contrast, this paper investigates whether beam-level \ac{SSB}-\ac{RSRP}, and \ac{CSI-RS}-\ac{RSRP} measurements,
can be used to infer the average downlink \ac{SINR} experienced over scheduled \acp{PRB}.
The rationale behind this approach is that \ac{RSRP} measurements implicitly encode the \ac{UE} geometry relative to the surrounding network, together with the effects of propagation, beamforming, and other channel characteristics. Since both the received signal power and the dominant interference components are shaped by these factors, a learning model can infer the latent relationship between the reported measurements and the resulting \ac{SINR}.

All simulations are conducted using \texttt{Giulia}\footnote{\texttt{Giulia} is publicly available at \url{https://github.com/giulia-open-lab/OpenGiuliaSLS.git}.}, a high-fidelity system-level simulator purpose-built for evaluating heterogeneous cellular networks through a combination of expert, \ac{3GPP}, and AI-based models~\cite{FR_mag_david}. 
Overall, these settings allow us to evaluate the feasibility of the proposed approach under widely adopted channel models, beam configuration, and scheduling conditions,
as a first step toward more scenario-specific predictive \ac{UE}-performance assessment involving real measurement data, online learning, explicit handover decisions, or dynamic cell switch-off procedures.

The main contributions can be summarized as follows:
\begin{itemize}
\item
We formulate \ac{UE} downlink \ac{SINR} estimation from \ac{5G} \ac{NR} reference-signal measurements as a supervised learning problem.

\item
We build a \ac{3GPP}-compliant synthetic dataset including \ac{SSB}-\ac{RSRP}, \ac{CSI-RS}-\ac{RSRP}, \ac{CSI-RS} beam activity information, and \ac{PRB}-level \ac{SINR} values in a multi-cell \ac{mMIMO} deployment.
\item
We design and compare different input feature representations for the \ac{SINR} predictor, including unfiltered beam measurements, Top-K strongest measurements, and Top-K active \ac{CSI-RS} measurements.
\item
We show that \ac{CSI-RS} beam-activity-aware feature selection significantly improves prediction accuracy,
while reducing the dimensionality of the input representation.
\end{itemize}

The remainder of this paper is organized as follows.
Section~\ref{sec:system_model} introduces the considered \ac{5G} \ac{NR} system model,
Section~\ref{sec:problem_formulation} formulates the indirect \ac{SINR} prediction problem,
and describes the input feature representations used by the learning model.
Section~\ref{sec:numerical_results} presents the simulation setup and discusses numerical results.
Finally,
Section~\ref{sec:conclusions} concludes the paper and outlines future research directions.
\section{System Model}
\label{sec:system_model}
In this section, we introduce the models used in this work to build our dataset and conduct our analysis. 
Here, we consider a 5G \ac{NR} cellular network operating in the sub-6\,GHz spectrum (\ac{FR1}) with carrier frequency $f_c = 3.5$\,GHz and full frequency reuse, for which the adopted models are outlined by \ac{3GPP}~\cite{3GPP38901,3GPP38214,3GPP38211}.
The network operates over a channel bandwidth of 100\,MHz, which, for a subcarrier spacing of 30\,kHz, corresponds to  $N_{\rm PRB} = 273$ \acp{PRB}, according to \ac{3GPP} specifications.

\subsubsection{Network and User Layout}
\begin{figure}[t!]
    \centering
    \includegraphics[width=0.45\textwidth]{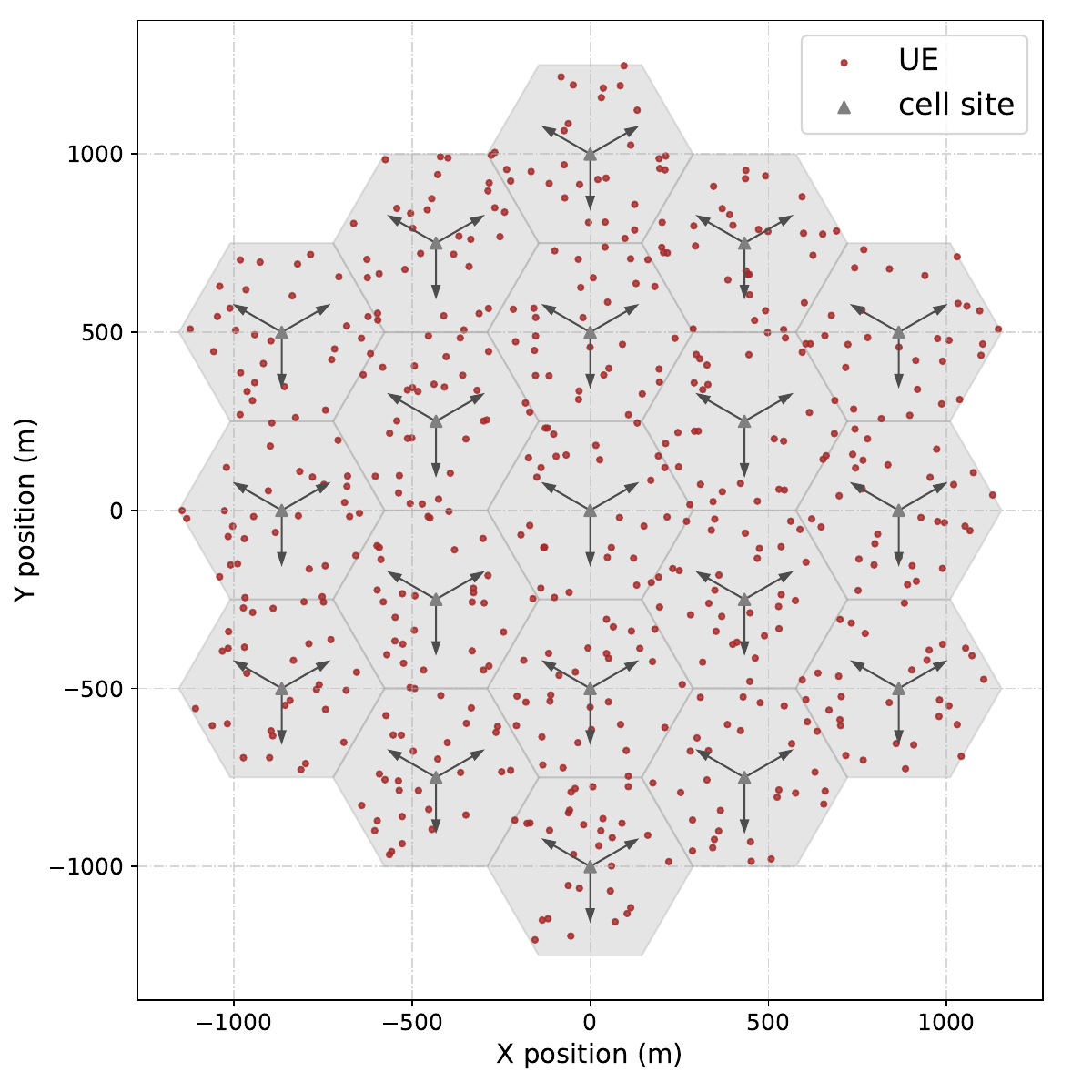}
    \vspace{-1em}
    \caption{2D Network Layout}
    \label{fig:2DNetworkLayout}
    \vspace{-2em}
\end{figure}

In our analysis, we consider an outdoor urban network comprising 19 sites arranged in a 2-tier hexagonal grid, with an inter-site distance $d_{\rm ISD}$. 
Each site has a height $h_{\rm BS}$ and hosts 3 sectors, whose boresight orientations are evenly spaced by 120$^\circ$.
The set of cells is represented as $\mathcal{C}$, with cardinality denoted as $N_{\rm c}$.
Here, we assume a fully loaded scenario with a total of $N_{\rm u}$ single-antenna \acp{UE} uniformly distributed over the network area, and we denote the set of all active \acp{UE} by $\mathcal{U}$.
Figure~\ref{fig:2DNetworkLayout} depicts an example of the resulting 2D network layout.

\subsubsection{Cell Antenna Array Unit}
Each 5G cell uses a \ac{UPA} with $M=32$ vertically polarized elements arranged in $M_v=4$ rows and $M_h=8$ columns, each connected to its own transceiver. The inter-element spacing is $\lambda_p/2$, with $\lambda_p$ equal to the carrier wavelength, and each \ac{UE} is equipped with a single antenna. The 5G cells leverage digital precoding and \ac{mMIMO} capabilities through active antenna arrays.
For each cell $c$, the matrix ${\bf V}_c = [{\bf v}_c^x,\,{\bf v}_c^y,\,{\bf v}_c^z]^T$ collects the Cartesian coordinates of its antenna elements w.r.t. the panel center~\cite{3GPP38901}.

\begin{figure*}[!tb]
\begin{equation}
\label{eq:SINR_comp}
\tag{7}
\gamma_{u,k} = 
    \frac{
    \beta_{u, \hat{c}_u} \left| {\bf h}_{u, \hat{c}_u, k}  {\bf w}_{u, \hat{c}_u, k} \right|^2 P_{u,\hat{c}_u, k}
    }
    {
    \beta_{u, \hat{c}_u}  \sum_{p \in \mathcal{U}_{\hat{c}_u} \setminus u}
     \left| {\bf h}_{u, \hat{c}_u, k}  {\bf w}_{p, \hat{c}_u, k} \right|^2   P_{p,\hat{c}_u, k} + 
    \sum_{b \in \mathcal{C} \setminus  \hat{c}_u} 
    \beta_{u, c}
    \sum_{i \in \mathcal{U}_c}
    \left| {\bf h}_{u, c, k}  {\bf w}_{i, c, k}\right|^2  P_{i,c,k} 
    +   
    \sigma_k^2
    }  
\vspace{0.5em}
\end{equation}
\hrulefill
\vspace{-1em}
\end{figure*}

\subsubsection{Channel Model}
We adopt the statistical \ac{UMa} channel model and single-antenna element gain specified by \ac{3GPP} in~\cite{3GPP38901}. This model characterizes the propagation between each \ac{UE} $u \in \mathcal{U}$ and each cell $c \in \mathcal{C}$ through large-scale and small-scale fading effects.

\textbf{Large-Scale Fading:}
For each \ac{UE} $u$ and cell $c$, the \ac{LoS} probability $P^{\rm LoS}_{u,c}$ and the path-loss gain $\rho_{u,c}$ follow the \ac{3GPP} \ac{UMa} models in~\cite{3GPP38901}. The shadow-fading gain $\tau_{u,c}$ is modeled as a zero-mean log-normal random variable whose standard deviation $\sigma_{\rm SF}$ depends on the link condition, with $\sigma_{\rm SF}=4$\,dB and $\sigma_{\rm SF}=6$\,dB for \ac{LoS} and \ac{NLoS} links, respectively~\cite{3GPP38901}. To reproduce realistic large-scale variations across the network, we further incorporate 2D spatial correlation among \acp{UE} following the model in~\cite{ShadowCorrelation_Xiaodong}.

The resulting large-scale gain between \ac{UE} $u$ and cell $c$ is
\begin{equation}\label{eq:LargeScaleCoeff}
\beta_{u,c}=g_{u,c}\,\rho_{u,c}\,\tau_{u,c}\;,
\end{equation}
where $g_{u,c}$ denotes the single-element antenna gain.

\textbf{Small-Scale Fading:}
The small-scale fading captures the multi-path nature of the link between each \ac{UE} $u$ and the $M$ antenna elements of cell $c$, and is modeled as a Rician channel~\cite{3GPP38901}. 
For each \ac{PRB} $k$, the downlink channel vector ${\bf h}_{u,c,k} \in \mathbb{C}^{1 \times M}$ is expressed as the weighted sum of a deterministic \ac{LoS} component and a stochastic \ac{NLoS} component,
\begin{equation}\label{eq:ComplexChannelRician}
    {\bf h}_{u,c,k} =
    \sqrt{\frac{K}{1+K}} \, {\bf h}^{\rm LoS}_{u,c,k}
    +
    \sqrt{\frac{1}{1+K}} \, {\bf h}^{\rm NLoS}_{u,c,k}\;,
\end{equation}
where the two terms are weighted by the Rician factor $K$, whose values are set according to~\cite{3GPP38901}. The \ac{NLoS} component is drawn independently for each \ac{PRB} as ${\bf h}^{\rm NLoS}_{u,c,k} \sim \mathcal{CN}({\bf 0}, {\bf I}_M)$. Embracing the plane-wave approximation~\cite{3GPP38901, massivemimobook}, the \ac{LoS} component models the phase shift between \ac{UE} $u$ and each of the $M$ antennas of cell $c$, and is given by
\begin{equation}\label{eq:ComplexChannelLOS}
    {\bf h}^{\rm LoS}_{u,c,k} = e^{-j \frac{2\pi}{\lambda_c} d^{\rm 3D}_{u,c}}
    \;
    e^{j \frac{2\pi}{\lambda_c} \, {\bf k}_{u,c}^T(\phi_{u,c}, \theta_{u,c}) \, {\bf V}_c}\;,
\end{equation}
where $d^{\rm 3D}_{u,c}$ is the 3D distance between \ac{UE} $u$ and the panel center of cell $c$, $\phi_{u,c}$ and $\theta_{u,c}$ are the corresponding azimuth and zenith angles, and we recall that ${\bf V}_c$ are the Cartesian coordinates of the panel antennas of cell $c$.
The wave vector ${\bf k}_{u,c}(\cdot, \cdot)$ represents the phase variation of a plane wave along the three orthogonal spatial directions and is defined as
\begin{equation}\label{eq:waveVector}
    {\bf k}_{u,c}(\phi_{u,c}, \theta_{u,c})
    =
    \begin{bmatrix}
    \cos(\phi_{u,c}) \cos(\theta_{u,c}) \\
    \sin(\phi_{u,c}) \cos(\theta_{u,c}) \\
    \sin(\theta_{u,c})
    \end{bmatrix}.
\end{equation}

\subsection{NR Initial Access}\label{subsec:NR_EnhInitialAcess}
During the initial cell-discovery and access phase, each cell $c$ transmits a set of \acp{RS} that enable each \ac{UE} to synchronize with the network and perform the measurements to select its serving cell and establish a connection for data transmission~\cite{DahlmanBook}. In \ac{NR}, these signals are jointly carried by the \ac{SSB}; hereafter, we refer to the beam used to transmit an \ac{SSB} as an \ac{SSB} beam. 
A distinctive feature of \ac{NR}, compared with previous cellular generations, is the ability to beamform multiple \ac{SSB} along multiple spatial directions, thereby improving signal strength, coverage flexibility, and interference management. According to the \ac{3GPP} specifications~\cite{3GPP38214, 3GPP38211}, each cell $c$ can transmit up to $N_c^{\rm ssb}$ \ac{SSB} beams, with $N_c^{\rm ssb} = 8$ in the \ac{FR1} band (i.e., sub-6\,GHz band). It should be noted that these beams are not transmitted simultaneously; instead, each cell performs a beam-sweeping procedure, transmitting them sequentially over time according to a predefined pattern.
Each \ac{SSB} beam is represented by a complex codeword ${\bf w}^{\rm ssb}_{s,c} \in \mathbb{C}^{M \times 1}$ drawn from a fixed codebook $\mathcal{W}_{\rm CB}^{\rm ssb}$ generated via \ac{2D-DFT}. 
Then, we denote with $\mathcal{W}^{\rm ssb}$ the set of the $N_c^{\rm ssb}$ transmitted \ac{SSB} beams.

Each \ac{UE} $u$ selects its serving cell by measuring the \ac{RSRP} of each transmitted \ac{SSB} beam. The \ac{RSRP} of beam $s$ from cell $c$ is given by
\begin{equation}\label{eq:rsrp_5G}
    {\rm RSRP}^{\rm ssb}_{u,s,c} = 
    \mathbb{E}_k \left[
    \beta_{u,c} \left| {\bf h}_{u,c,k} {\bf w}_{s,c}^{\rm ssb} \right|^2 P^{\rm ssb}_{s,c}
    \right] \;,
\end{equation}
where we recall that $P^{\rm ssb}_{s,c}$ is the \ac{SSB} beam transmission power.
Here, the expectation is taken over multiple channel realizations across the resource elements assigned to reference signals.
Finally, each \ac{UE} $u$ selects its serving cell $\hat{c}_u$ as the one yielding the highest \ac{RSRP}.

\subsection{Data Precoding and PRB SINR}
Each cell $c$ serves its associated \acp{UE} $\mathcal{U}_c$ by multiplexing the transmitted data across \acp{PRB} and beams.
To exploit the \ac{mMIMO} beamforming and multiplexing capabilities of the 5G cells, we adopt a Type-I \ac{CSI-RS} Codebook precoding scheme~\cite{3GPP38214, DahlmanBook}.
According to this scheme, to define the data precoding of each \ac{UE} per \ac{PRB}, each cell transmits a set of \acp{CSI-RS} for the channel estimation phase to its set of connected \acp{UE} $\mathcal{U}_c$.
Analogously to the \ac{SSB} beams, each \ac{CSI-RS} beam is represented by a complex codeword ${\bf w}^{\rm csi}_{s,c} \in \mathbb{C}^{M \times 1}$ drawn from a fixed codebook $\mathcal{W}_{\rm CB}^{\rm csi}$ generated via \ac{2D-DFT} and mapped to specific resource elements\footnote{To simplify the notation, the superscript `csi' is used in place of `CSI-RS' throughout the paper.}.
In accordance with the \ac{3GPP} \ac{FR1} specifications, each cell can transmit up to $N_c^{\rm csi}$ \ac{CSI-RS} beams, with $N_c^{\rm csi} = 32$; we then denote with $\mathcal{W}^{\rm csi}$ the set of the $N_c^{\rm csi}$ transmitted \ac{CSI-RS} beams.
Then, each \ac{UE} estimates the received power of each \ac{CSI-RS} beam to feed back a set of indices characterizing the instantaneous channel conditions. Specifically, the \ac{CSI-RS} \ac{RSRP} of beam $s$ from cell $c$ over \ac{PRB} $k$ is computed as follows,
\begin{equation}\label{eq:rsrp_csi}
    {\rm RSRP}^{\rm csi}_{u,s,c,k} = 
    \mathbb{E}\left[
    \beta_{u,c} \left| {\bf h}_{u,c,k}\, {\bf w}_{s,c}^{\rm csi} \right|^2 P^{\rm csi}_{s,c}
    \right] \;,
\end{equation}
where $P^{\rm csi}_{s,c}$ is the \ac{CSI-RS} beam transmission power.
Unlike the \ac{SSB}, however, the \ac{CSI-RS} is mapped to a dedicated set of resource elements and is resolved per \ac{PRB}, so as to capture the short-term per-\ac{PRB} conditions that drive the data precoding.
Among the fed-back indices, the \ac{PMI} indicates the beam yielding the largest received power, and it will be used by the serving cell to set the precoding vector ${\bf w}_{u, \hat{c}_u, k}$ of \ac{UE} $u$ equal to the precoder associated with that beam.

Within each cell, \acp{PRB} are fully reused across beams and equally split among the \acp{UE} served per beam, thereby assuming equal long-term resource allocation among \acp{UE}, as typically done in round-robin schedulers.
Additionally, it should be noted that in this work we consider a single-layer transmission; therefore, a \ac{UE} is served by only one beam. 

Letting ${\bf w}_{u,c,k}$ denote the precoding codeword assigned to \ac{UE} $u$ by cell $c$ on \ac{PRB} $k$ and $P_{u,c,k}$ the corresponding transmit power, the data \ac{SINR} per \ac{PRB} $\gamma_{u,k}$ is given by eq.~\eqref{eq:SINR_comp}.
Here, the numerator represents the useful received power, i.e., the power received by \ac{UE} $u$ from the selected beam on resource $k$. In the denominator, the first term accounts for intra-cell inter-beam interference, namely the power received on \ac{PRB} $k$ from other beams of the serving cell sharing the same resource; the second term represents inter-cell interference from neighboring cells. Finally, $\sigma^2_k$ corresponds to the receiver noise power.
\setcounter{equation}{7}
\section{Problem Formulation}
\label{sec:problem_formulation}
In this work, we aim at estimating for a given user $u$ the average \ac{SINR} across all its scheduled \acp{PRB}, denoted as $\overline{\gamma_u}$, from beam-level \ac{RSRP} reports. Specifically, the report here comprises the \ac{RSRP} measured on the \ac{SSB} beams, ${\rm RSRP}_{u,s,c}^{\rm ssb}$, and on the \ac{CSI-RS} beams, ${\rm RSRP}_{u,s,c}^{\rm csi}$, of the serving cell $\hat{c}_u$ and the neighbouring cells $c$.

As detailed in Section~\ref{sec:system_model}, user data are precoded using Type-I codebook-based feedback~\cite{3GPP38214}. However, this feedback exposes only the dominant beam of the serving cell and is therefore insufficient to characterize the multi-cell, multi-beam interference that shapes the \ac{SINR}.
We instead build on the \ac{NR} beam-management framework, in which each \ac{UE} measures the Layer-1 \ac{RSRP} on its configured \ac{SSB} and \ac{CSI-RS} resources (as defined in~\cite{3GPP38215}) and reports them to the network with the corresponding resource indicators~\cite{3GPP38214}.
While practical deployments typically report only a subset of these measurements~\cite{3GPP38215}, we assume a setting in which the network has access to the complete per-beam measurement report of every \ac{UE}. Consequently, ${\rm RSRP}_{u,s,c}^{\rm ssb}$ and ${\rm RSRP}_{u,s,c}^{\rm csi}$ are available for all serving and neighboring beams.

Let $f_\theta(\cdot)$ denote a parameterized estimator. Our objective is to determine the optimal model parameters $\theta^*$ that minimizes the \ac{MSE} between the estimated and target average \ac{SINR} across all \acp{UE} $u\in\mathcal{U}$:
\begin{equation}
\theta^* = \argmin_\theta \frac{1}{|\mathcal{U}|} \sum_{u\in\mathcal{U}} \left| \overline{\gamma}_u - f_\theta({\bf x}_u) \right|^2,
\end{equation}
where ${\mathbf{x}}_u$ is the input feature vector associated with \ac{UE} $u$. 

\subsection{SINR Predictor Input Feature}
The construction of the input feature vector ${\mathbf{x}}_u$ for the \ac{SINR} predictor is a key design choice. In the following, we introduce three alternative feature representations, namely: \textit{Unfiltered}, \textit{Top-K Strongest}, \textit{Top-K Active}.

\subsubsection{Unfiltered}
The baseline representation constructs ${\mathbf{x}}_u$ by concatenating all 
measurement reports from both the serving cell and neighboring cells using a fixed, predefined order for all users.
Let $\mathbf{r}_u^{\mathrm{ssb}} \in \mathbb{R}^{N_{\mathrm{ssb}}}$ and $\mathbf{r}_u^{\mathrm{csi}} \in \mathbb{R}^{N_{\mathrm{csi}}}$ denote the row vectors collecting all \ac{SSB}- and \ac{CSI-RS}-based \ac{RSRP} measurements of user $u$, with $N_{\mathrm{ssb}}=N_c N_c^{\mathrm{ssb}}$ and $N_{\mathrm{csi}}=N_c N_c^{\mathrm{csi}}$. Then the input feature is simply defined as:
\begin{equation}
    \mathbf{x}_u = [\mathbf{r}_u^{\mathrm{ssb}} \quad \mathbf{r}_u^{\mathrm{csi}}]^{\mathrm{T}}
\end{equation}

\subsubsection{Top-K Strongest}
A second feature engineering strategy involves sorting the measurement reports by their \ac{RSRP} magnitudes. 
To build this, let $r_{u,(i)}^{\rm ssb}$ denote the $i$-th largest element of the vector ${\mathbf{r}}_u^{\rm ssb}$, such that $r_{u,(1)}^{\rm ssb} \ge r_{u,(2)}^{\rm ssb} \ge \dots \ge r_{u,(N_{\rm ssb})}^{\rm ssb}$.
Similarly, let $r_{u,(i)}^{\rm csi}$ denote the $i$-th largest element of the vector  ${\mathbf{r}}_u^{\rm csi}$.
Then, the
\textit{Top-K Strongest} user input feature is built by only considering the Top-$K_{\rm ssb}$ \ac{SSB} RSRP reports and the Top-$K_{\rm csi}$ \ac{CSI-RS} RSRP, equivalently:
\begin{equation}
    \mathbf{x}_u=[r_{u,(1)}^{\rm ssb}, \dots , r_{u,(K_{\rm ssb})}^{\rm ssb}, r_{u,(1)}^{\rm csi}, \dots , r_{u,(K_{csi})}^{\rm csi} ]^{\rm T}
\end{equation}
This filtering has twofold advantages: first, it reduces the input cardinality from 
$N_{\rm ssb} + N_{\rm csi}$ to $K_{\rm ssb} + K_{\rm csi}$, where both limits can be parameters subject to optimization; second, ordering the values ensures that the serving beams (typically the strongest) and the primary interfering beams are at the beginning of the vectors, mitigating the combinatorial complexity of the unfiltered approach, which relied on unstructured positional encoding

\subsubsection{Top-K Active}
However, a strong measurement from an interfering beam does not necessarily translate into realized interference. 
If the corresponding \ac{CSI-RS} beam is inactive, i.e., no data are transmitted via that beam across any \ac{PRB}, it does not generate interference. 
To account for this, we define a binary activity mask vector $\mathbf{a}^{\rm csi}$, whose elements are $1$ if the corresponding \ac{CSI-RS} beam is active and serving at least one \ac{UE}, and $0$ otherwise. 
Applying this mask to the \ac{CSI-RS} measurements yields the activity-filtered vector $\tilde{\mathbf{r}}_u^{\rm csi} = \mathbf{r}_u^{\rm csi} \odot \mathbf{a}^{\rm csi}$. 
Letting $\tilde{r}_{u,(i)}^{\rm csi}$ denote the $i$-th largest element of this masked vector, the \textit{Top-K Active} user feature is formulated as:
\begin{equation}
    \mathbf{x}_u = \left[ r_{u,(1)}^{\rm ssb}, \dots, r_{u,(K_{\rm ssb})}^{\rm ssb}, \tilde{r}_{u,(1)}^{\rm csi}, \dots, \tilde{r}_{u,(K_{\rm csi})}^{\rm csi} \right]^{\rm T}
\end{equation}
Notably, because we assume that data is scheduled in orthogonal resources to the \acp{SSB} beams,  
\ac{SSB} transmissions do not generate data-channel interference and are therefore excluded from the activity mask.
\section{Numerical results}
\label{sec:numerical_results}
In this section, we evaluate the numerical performance of the proposed \ac{SINR} predictor. 
The dataset used to train the estimator was generated using \texttt{Giulia}, a high-fidelity open-source system-level simulator, 
configured to follow the specification of the \ac{UMa} model reported in the \ac{3GPP} TR38.901 document~\cite{3GPP38901}.

\subsection{Synthetic Dataset Generation}
The dataset includes $N$ independent simulations, where each simulation captures the following main events:
\begin{itemize}
    \item $N_{\rm u}$ users are randomly and uniformly spread within the considered region;
    \item each \ac{UE} $u\in\mathcal{U}$ connects to the strongest cell $\hat{c}_u$ based on the strongest measured \ac{SSB}-\ac{RSRP};
    \item a beam refinement procedure detects the strongest \ac{CSI-RS} \ac{RSRP} from the serving cell, determining the beam index for data transmission, 
    i.e., the data precoding vector $\mathbf{w}_{u,\hat{c}_u,k}$ used to transmit data;
    \item Each serving cell schedules transmissions to its associated \acp{UE} across \acp{PRB} and beams, using the serving beam identified in the previous step.
\end{itemize}

For each simulation, and for each user, we record the following output:
\begin{itemize}
    \item The list of \ac{RSRP} measurements from all \ac{SSB} and \ac{CSI-RS} signals transmitted by all cells, yielding the vectors $\mathbf{r}_u^{\rm ssb}$ and $\mathbf{r}_{u}^{\rm csi}$;
    \item The set of \acp{PRB} scheduled to user $u$ from its serving cell $\hat{c}_u$, denoted as $\mathcal{K}_u$;
    \item The sequence of \ac{SINR} values $\gamma_{u,k}$ for each scheduled \ac{PRB} $k\in \mathcal{K}_u$;
    \item The activity mask vector $\mathbf{a}^{\rm csi}$ for the current simulation.
\end{itemize}
The target label $\overline{\gamma}_u$ is then calculated as the average \ac{SINR} over the scheduled \acp{PRB}:
\begin{equation}
    \overline{\gamma}_u = \frac{1}{|\mathcal{K}_u|}\sum_{k\in K_u} \gamma_{u,k}
\end{equation}

\subsection{SINR Predictor Training}
The generated dataset, comprising $N \cdot N_{\rm u}$ entries, was shuffled and partitioned into training, validation, and test sets using a 70/15/15 split. 
To prevent data leakage, all input features were standardized, i.e., scaled to zero mean and unit variance, using only the training set statistics; this identical transformation was then applied to the validation and test sets. 
The \ac{SINR} predictor is a three-layer \ac{MLP} parameterized by $\theta$, featuring two hidden layers with ReLU activation and a linear output layer.
It was trained via backpropagation to minimize the \ac{MSE} on mini-batches of size $Z$, utilizing the Adam optimizer with a learning rate of $\alpha$. 
To mitigate overfitting, training was capped at a maximum of 500 epochs, with early stopping enforced if the validation loss failed to decrease for 100 consecutive epochs. All simulation and training settings are summarized in Table~\ref{tab:parameters}.
\begin{table}[t]
    \centering
    \caption{Simulation Parameters}
    \label{tab:parameters}
    \footnotesize
    \begin{tabularx}{\linewidth}{@{}Y Y @{\hspace{2em}} Y Y@{}}
        \toprule
        \textbf{Parameter} & \textbf{Value} & \textbf{Parameter} & \textbf{Value} \\
        \midrule
        $N$              & $1500$           & $N_{\rm u}$                      & $570$       \\
        $N_{\rm c}$            & $57$             & $N_{\rm ssb}$                    & $456$ \\
        $N_{\rm csi}$ & $1824$           & $K$                              & $273$        \\
        $\theta$         & $258 \times 128 \times 1$ & $\alpha$                              & $0.001$     \\
        $Z$              & $2048$           & $[K_{\rm ssb},\, K_{\rm csi}]$ & $[32,\,16]$ \\
        \bottomrule
    \end{tabularx}
    \vspace{-1.5em}
\end{table}

\subsection{Results Discussion}

\begin{figure}[!t]
    \centering
    \hspace*{-0.5cm}
    \subfloat {
        \setlength\fwidth{0.975\columnwidth}
        \setlength\fheight{0.6026\columnwidth}
        \input{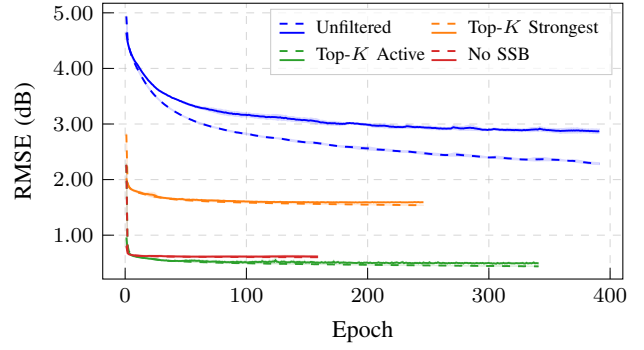}
    }
    \vspace{-1em}
    \caption{\ac{SINR} predictor input feature $\mathbf{x}_u$ comparison.\label{fig:feature_comparison}.
    Solid and dashed lines denote validation and training loss, respectively.}
    \vspace{-1.5em}
\end{figure}
Fig.~\ref{fig:feature_comparison} shows the learning performance, evaluated in terms of \ac{RMSE}, on the training and test sets over epochs for the three different input features defined in Sec.~\ref{sec:problem_formulation}. It also includes an ablation study on \textit{Top-K~Active} where we removed the top $K_{\rm ssb}$ strongest \ac{RSRP} vector, resulting in four evaluated configurations: \mbox{\textit{Unfiltered}}, \mbox{\textit{Top-K~Strongest}}, \mbox{\textit{Top-K~Active}}, and \mbox{\textit{No~SSB}}.
As expected, the \textit{Top-K~Active} configuration outperforms the other input feature alternatives. This is because the predictor receives an ordered sequence of \ac{RSRP} values for active beams, allowing it to infer the resulting user \ac{SINR}. In contrast, the \mbox{\textit{Top-K~Strongest}} approach yields lower precision because the model cannot determine which of the strong beams affect the \ac{UE} \ac{SINR}. However, this approach does not require the binary activity vector, and users can report the ordered measurements directly to the serving cell. The \textit{Unfiltered} input, i.e., the basic input feature with raw \ac{RSRP} measurements, fails to train an effective predictor. Because the input vector leverages a fixed order of beams, the \ac{SINR} prediction is forced to learn a combinatorial problem; it treats an input differently if interfering beam $i$ and beam $j$ are swapped, even if they contribute equally to the \ac{SINR} estimation.
Finally, the \mbox{\textit{No~SSB}} configuration further reduces the input feature size; however, by removing the \ac{RSRP} associated with the received \ac{SSB} signals, the predictor overfits earlier and performs worse. This suggests that even if \ac{SSB} beams do not actively contribute to the \ac{SINR} calculation, they enrich the estimation by providing valuable information.


\begin{figure}[!t]
    \centering
    \hspace*{-0.5cm}
    \subfloat {
        \setlength\fwidth{0.975\columnwidth}
        \setlength\fheight{0.6026\columnwidth}
        \input{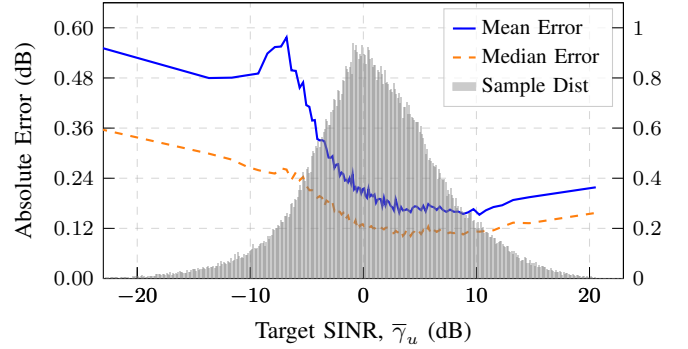}
    }
    \vspace{-2em}
    \caption{Absolute prediction error distribution of the proposed \textit{Top-K Active} predictor as a function of the target \ac{SINR}.}
    \label{fig:error_distribution} 
    \vspace{-1.5em}
\end{figure}
Regarding the \textit{Top-K~Active} \ac{SINR} predictor, Fig.~\ref{fig:error_distribution} reports the mean and median prediction errors in~dB as a function of the target \ac{SINR}, alongside the distribution of the true \ac{SINR} in the test dataset.
We observe that the \ac{SINR} follows a Gaussian distribution with a mean of approximately 1~dB and a standard deviation of about 6~dB, indicating that the \ac{3GPP} \ac{UMa} is a strong interference-limited scenario.
As expected, the \ac{SINR} prediction error is smaller where most of the samples are concentrated.
However, the absolute error distribution becomes asymmetrical at the tails, exhibiting higher errors in the lower \ac{SINR} regime compared to the high \ac{SINR} regime. Moreover, the growing difference between the median and mean absolute errors indicates the presence of strong outliers.


\begin{figure}[!t]
    \centering
    \hspace*{-0.5cm}
    \subfloat {
        \setlength\fwidth{0.975\columnwidth}
        \setlength\fheight{0.6026\columnwidth}
\begin{tikzpicture}

\definecolor{darkgray176}{RGB}{176,176,176}
\definecolor{gray}{RGB}{128,128,128}
\definecolor{darkred}{RGB}{139,0,0}

\begin{axis}[
width=\fwidth,
height=\fheight,
log basis x={2},
log ticks with fixed point,
tick align=inside,
tick pos=left,
x grid style={darkgray176, opacity=0.5, dashed},
xlabel={Top Active CSI-RS RSRP, \(\displaystyle K_{\rm csi}\)},
xlabel style={font=\small},
xmajorgrids,
xmin=6.49801917084988, xmax=630.345939632597,
xmode=log,
xtick style={color=black},
xticklabel style={font=\footnotesize},
x grid style={darkgray176, opacity=0.5, dashed},
ylabel={Test RMSE (dB)},
ylabel style={font=\small},
ymajorgrids,
ymin=0.474803930174827, ymax=0.557913376973772,
ytick={0.48,0.49,0.50,0.51,0.52,0.53,0.54,0.55},
ytick style={color=black},
y grid style={darkgray176, opacity=0.5, dashed},
yticklabel style={
    font=\footnotesize,
    /pgf/number format/fixed,
    /pgf/number format/fixed zerofill,
    /pgf/number format/precision=2},
]
\path [fill=blue, fill opacity=0.25]
(axis cs:8,0.502841586053578)
--(axis cs:8,0.496584503267294)
--(axis cs:16,0.478354359574779)
--(axis cs:32,0.481036017175756)
--(axis cs:64,0.488088657121187)
--(axis cs:128,0.495686221334164)
--(axis cs:256,0.508625499716565)
--(axis cs:512,0.540612170377801)
--(axis cs:512,0.54936294757382)
--(axis cs:512,0.54936294757382)
--(axis cs:256,0.520484177601671)
--(axis cs:128,0.503594840601906)
--(axis cs:64,0.499731635514669)
--(axis cs:32,0.486987838478897)
--(axis cs:16,0.485799425715903)
--(axis cs:8,0.502841586053578)
--cycle;

\addplot [semithick, blue, mark=*, mark size=1.5, mark options={solid}]
table {%
8 0.499713044660436
16 0.482076892645341
32 0.484011927827326
64 0.493910146317928
128 0.499640530968035
256 0.514554838659118
512 0.54498755897581
};
\addplot [thick, darkred, dashed]
table {%
16 0.474803930174827
16 0.557913376973772
};
\end{axis}

\end{tikzpicture}
    }
    \vspace{-1em}
    \caption{$K_{\rm csi}$ comparison.\label{fig:K_CSI}}
    \vspace{-1em}
\end{figure}
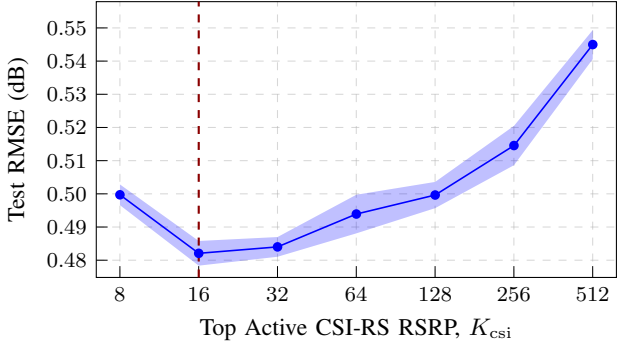
As highlighted by the previous results, the Top-$K$ selection of active beams is a fundamental feature engineering step that can drastically improve the \ac{SINR} prediction. However, determining which beams transport data, potentially at each scheduling time period, requires communication overhead between cells.
Fig.~\ref{fig:K_CSI} reports the test set \ac{RMSE} as a function of the number of ordered top active \ac{CSI-RS} \ac{RSRP} measurements provided as input to the predictor, denoted as $K_{\rm csi}$. Notably, an optimal value of $\hat{K}_{\rm csi}=16$ jointly minimizes the estimation error and reduces the inter-cell communication overhead.
When $K_{\rm csi}<\hat{K}_{\rm csi}$, too few active interferers are considered, limiting the \ac{SINR} prediction accuracy. Conversely, when $K_{\rm csi}>\hat{K}_{\rm csi}$, the input features include more interfering beam \ac{RSRP}s, but their contribution to the final \ac{SINR} estimation is less significant. This yields diminishing returns, considering the additional inter-cell information exchange required to support the estimation.


\begin{figure}[!t]
    \centering
    \hspace*{-0.5cm}
    \subfloat {
        \setlength\fwidth{0.975\columnwidth}
        \setlength\fheight{0.6026\columnwidth}
\begin{tikzpicture}

\definecolor{darkgray176}{RGB}{176,176,176}
\definecolor{gray}{RGB}{128,128,128}
\definecolor{darkred}{RGB}{139,0,0}

\begin{axis}[
width=\fwidth,
height=\fheight,
log basis x={2},
log ticks with fixed point,
tick align=inside,
tick pos=left,
x grid style={darkgray176, opacity=0.5, dashed},
xlabel={Top Strongest SSB RSRP, \(\displaystyle K_{\rm ssb}\)},
xlabel style={font=\small},
xmajorgrids,
xmin=6.49801917084988, xmax=630.345939632597,
xmode=log,
xtick style={color=black},
xticklabel style={font=\footnotesize},
x grid style={darkgray176, opacity=0.5, dashed},
ylabel={Test RMSE (dB)},
ylabel style={font=\small},
ymajorgrids,
ymin=1.93489024495897, ymax=2.53122444964222,
ytick={2.00, 2.10, ..., 2.6},
ytick style={color=black},
y grid style={darkgray176, opacity=0.5, dashed},
yticklabel style={
    font=\footnotesize,
    /pgf/number format/fixed,
    /pgf/number format/fixed zerofill,
    /pgf/number format/precision=2},
]

\path [fill=blue, fill opacity=0.25]
(axis cs:8,2.50411834942935)
--(axis cs:8,2.48768450144492)
--(axis cs:16,2.02181144485195)
--(axis cs:32,1.96199634517185)
--(axis cs:64,1.96554926233758)
--(axis cs:128,1.97348465499466)
--(axis cs:256,1.99380155971621)
--(axis cs:456,2.01227659159472)
--(axis cs:456,2.02867155824234)
--(axis cs:456,2.02867155824234)
--(axis cs:256,2.05778776233694)
--(axis cs:128,2.00060032875116)
--(axis cs:64,2.00064438246712)
--(axis cs:32,1.9818439452121)
--(axis cs:16,2.03064326623235)
--(axis cs:8,2.50411834942935)
--cycle;

\addplot [thick, blue, mark=*, mark size=1.5, mark options={solid}]
table {%
8 2.49590142543713
16 2.02622735554215
32 1.97192014519198
64 1.98309682240235
128 1.98704249187291
256 2.02579466102657
456 2.02047407491853
};
\addplot [semithick, darkred, dashed]
table {%
32 1.93489024495897
32 2.53122444964222
};
\end{axis}

\end{tikzpicture}
    }
    \vspace{-1em}
    \caption{$K_{\rm ssb}$ comparison.\label{fig:K_SSB}}
    \vspace{-1.5em}
\end{figure}
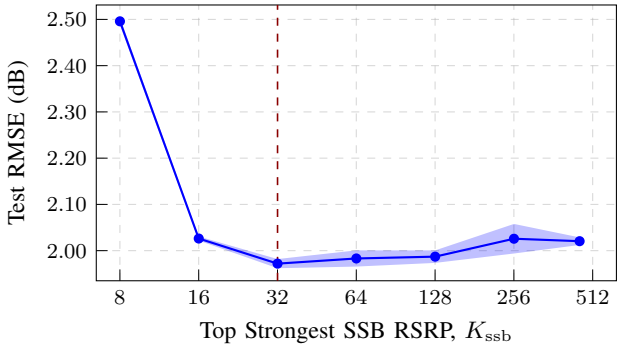
Similarly, Fig.~\ref{fig:K_SSB} shows the sweep search for the optimal value of $K_{\rm ssb}$ to use as an input feature. For this experiment, we consider an input feature set containing only the \ac{RSRP} reports from \ac{SSB} beams, $\mathbf{r}_u^{\rm ssb}$, omitting the \ac{CSI-RS} reports.
By relying solely on \ac{SSB} reports, the \ac{SINR} prediction error ranges up to 2.9~dB, achieving its minimum of 2.4~dB when $\hat{K}_{\rm ssb}=32$. This is because data is transmitted via \ac{CSI-RS} beams; therefore, they play a more prominent role in the final \ac{SINR} prediction.
As $K_{\rm ssb}$ decreases, the spatial signature of user $u$ becomes less precise, leading to significant losses in \ac{SINR} prediction accuracy. While higher values of $K_{\rm ssb}$ do not degrade performance, the increased number of measurements each \ac{UE} must perform and feed back to the serving cell yields diminishing returns.
\section{Conclusions}
\label{sec:conclusions}
In this paper, we introduced a supervised learning framework to predict the average downlink \ac{SINR} of \acp{UE} in 5G \ac{NR} networks leveraging standardized \ac{SSB} and {CSI-RS} \ac{RSRP} measurements report.
Our comparative analysis of various input feature representations revealed that a \textit{Top-K Active} filtering strategy, which accounts for actual \ac{CSI-RS} activity, significantly outperforms unfiltered or strictly strength-based approaches.
Furthermore, we demonstrated the existence of an optimal input dimensionality \mbox{($K_{\rm csi}=16$ and $K_{\rm ssb}=32$)} that minimizes estimation error while limiting the necessary inter-cell communication overhead. 
Ultimately, this data-driven methodology proves highly effective for proactive radio-metric estimation. Future research will focus on extending this predictive framework to real-world measurement datasets and utilizing digital twin sandboxing to simulate and optimize closed-loop network control policies, such as predictive handover and dynamic energy-saving procedures.
\section*{Acknowledgment}
\vspace{-0.05cm}
This research is supported by the action CNS2023-144333, financed by MCIN/AEI/10.13039/501100011033 and the European Union “NextGenerationEU”/PRTR.
\bibliographystyle{IEEEtran}
\bibliography{IEEEabrv, main.bib}
\end{document}